\documentclass[11pt]{article}
\usepackage{epsfig}
\usepackage{latexsym}
\usepackage{amsmath}

\renewcommand{\forall}{\mbox{for all}\,\,}
          \def\dt{\cal}
          \def\dA{{\dt A}}
          \def\dB{{\dt B}}

          \def\dH{{\dt H}}

          \def\B{{\cal B}}
          \def\C{{\cal C}}

          \def\H{{\cal H}}

          \def\K{{\cal K}}
          
          \def\M{{\cal M}}
          
          \def\O{{\cal O}}
          \def\P{{\cal P}}

          \def\W{{\cal W}}
          \def\X{{\cal X}}

          \def\gO{\Omega}

          \def\eps{\varepsilon}
          
          \def\gg{\gamma}

          \def\gs{\sigma}

          \def\aloc{\dA_{\rm loc}}
          
          \def\complex{{\bf C}}
          \def\del{\partial}

          \def\Halmos{\quad\hfill$\Box$}
          \def\id{\rm id}

          \def\pct{P$_1$CT}

          \def\Rd{\reals^{1+s}}
          \def\reals{{\bf R}}
          
          \def\supp{\rm supp}
          
\title{Localization Regions of Local Observables}
\author{Bernd Kuckert\thanks{Casimir-Ziegler fellow of the
Nordrhein-Westf\"alische Akademie der Wissenschaften}\\
Universiteit van Amsterdam\\
Korteweg-de Vries Instituut voor Wiskunde\\Plantage Muidergracht 24
1018 TV Amsterdam, Netherlands\\e-mail:kuckert@wins.uva.nl}
\date{}

\begin{document}

\frenchspacing
\maketitle

\begin{abstract}
Exploiting the properties of the Jost-Lehmann-Dyson representation, it
is shown that in 1+2 or more spacetime dimensions, a nonempty smallest
localization region can be associated with each local observable
(except for the c-numbers) in a theory of local observables in the
sense of Araki, Haag, and Kastler. Necessary and sufficient 
conditions are given that observables with spacelike separated localization 
regions commute (locality of the net alone does not imply this yet).
\end{abstract}

\section{Introduction}

The algebraic approach to relativistic quantum physics \cite{Haa92}
aims at joining the structures familiar from nonrelativistic quantum
mechanics to those of special relativity.  The object of investigation
is a net $\dA$ of {\em local observables} that associates with every
bounded open region $\O$ in the Minkowski spacetime $\Rd$ a unital
C$^*$-algebra $\dA(\O)$ of bounded operators in a Hilbert space $\H$
in such a way that $\O\subset P\subset\Rd$ implies
$\dA(\O)\subset\dA(P)$ ({\em isotony}), and such that the elements of
algebras associated with spacelike separated regions commute ({\em
locality}). For every region $\O$, the elements of $\dA(\O)$ are
interpreted as the observables measurable in a lab located in $\O$.

This paper deals with the question whether, given a single element
$A\in\dA(\O)$ for some bounded $\O\subset\Rd$, one can find a smallest
nonempty region $L(A)$ within which $A$ can be measured. From a
theorem of Landau it is well known that in at least 1+2 spacetime
dimensions, one observable cannot be measured in two labs located in
disjoint spacetime regions. A generalization of this theorem will be
proved below, and for at least 1+2 spacetime dimensions, it will lead
to several meaningful definitions of convex localization regions. Some
additional technical assumptions then provide a strongest localization
prescription with the property that observables with spacelike
separated localization regions commute.  But the fact that locality of
the net alone (not even together with those assumptions that lead to
the definition of a nonempty localization region) does not imply this
version of locality, is a little surprising.

This article is structured as follows: Section \ref{notation}
discusses the notation, concepts, and basic assumptions that play a
role in this paper. Section \ref{wave equation techniques} collects
and completes the tools that will be used in what follows. Section
\ref{results} contains the main results of the paper. In Section
\ref{empty-intersection} the theorem due to Landau mentioned above is
discussed and generalized: it states that in 1+2 or more spacetime
dimensions, the algebras of any two double cones with disjoint
closures only have in common the complex multiples of the identity
operator.  Using the techniques introduced in the first sections, it
is shown that this result can be generalized to the {\em
empty-intersection theorem}: instead of considering two double cones,
one can consider one double cone and any finite number of wedge
regions; if the closures of the regions under consideration have an
empty common intersection, then the corresponding algebras have in
common exactly the c-numbers.

In Section \ref{nonempty-intersection}, finally, it is discussed how
one can, in 1+2 or more spacetime dimensions, use the
empty-intersection theorem in order to associate a nonempty {\em
localization region} with any given {\em single} local observable that
is not a multiple of the identity.  But even though locality of the
net is assumed from the outset, the question whether local observables
with spacelike separated localization regions commute turns out to be
nontrivial.  A necessary and sufficient criterion for the locality of
such a localization prescription is provided by the {\em
nonempty-intersection theorem}: the criterion requires that, given any
finite family of wedges, all local observables contained in the
algebras associated with all these wedges are contained in the algebra
associated with any neighbourhood of the intersection of the wedges as
well. It is shown that an additional additivity assumption, which is
typically fulfilled by nets arising from Wightman fields, implies this
criterion, so it should be quite difficult to find nonpathological
counterexamples. On the other hand it is illustrated why the
nonempty-intersection criterion, though looking quite natural, is far
from self-evident. It does not follow from the locality property of
the net.

In the Conclusion, some related results are discussed briefly.

\section{Preliminaries}

\subsection{Notation and assumptions}\label{notation}

In what follows, $\H$ will be an infinite-dimensional (not necessarily separable)
Hilbert space, and $\dA$ will be a
net of observables as defined above, i.e., a net which satisfies
locality and isotony. The union of all algebras
$\dA(\O)$ associated with bounded open regions\footnote{In this article we 
refer to arbitrary subsets of $\Rd$ as \lq regions\rq.} $\O\subset\Rd$,
$s\geq1$, is an
involutive algebra $\aloc$ called the algebra of {\em local observables}.
Throughout Section \ref{results} it will be assumed that $s\geq2$, but for the 
introductory part this assumption is not yet necessary.

The following (standard) assumptions on $\dA$ will be made throughout this 
article:

\begin{quote}
{\bf (A) Translation covariance.} $\dA$ is
{\em covariant} under a strongly continuous unitary
representation $U$ of the group $(\Rd,+)$ of spacetime translations, 
i.e., 
$$U(a)\dA(\O)U(-a)=\dA(\O+a)$$
for every bounded region $\O$ and every $a\in\Rd$.

{\bf (B) Spectrum condition.} The spectrum of the four-momentum operator
generating $U$ is contained in the closure of the forward light cone.

{\bf (C) Existence and uniqueness of the vacuum.} 
The space of $U$-invariant vectors in $\H$ is
one-dimensional. $\gO$ will
denote an arbitrary, but fixed unit vector $\gO$ in this space, 
the {\em vacuum vector}. $\gO$ is {\em cyclic} with respect to 
$\aloc$, i.e., $\overline{\aloc\gO}=\H$.
\end{quote}
Throughout Section \ref{results} the following additional assumption will be made:
\begin{quote}
{\bf (D) Reeh-Schlieder property.} For every nonempty bounded 
open region $\O$, the space
$\dA(\O)\gO$ is dense in $\dH$.
\end{quote}
Assumptions (A) and (B) make sure that the system described by the net
has a well-defined
four-momentum whose spectrum ensures energetic stability of the system.

Given Assumptions (A) and (B), Assumption (C) holds if and only if 
$\gO$ induces a unique and pure vacuum state, which, in turn, holds
if and only if the algebra $\aloc$ is irreducible. A sufficient 
condition for this uniqueness is that
the bicommutant $\aloc''$ of $\aloc$ is a factor 
(Thm. III.3.2.6 in \cite{Haa92}). But as soon as $\H$ is separable,
this implies that the uniqueness
part of Assumption (C) does not mean any loss of generality: 
every von Neumann algebra in a separable Hilbert space
admits a direct-integral decomposition
into factors, and since the unitaries representing the translations
commute with the elements of the center of $\aloc''$, one can conclude that
almost all factors of the central
decomposition inherit Properties (A) and (B) (cf. also the remarks in
\cite{Haa92}, Sect. III.3.2, and references therein).

The Reeh-Schlieder property, Assumption (D), 
holds for all Wightman fields \cite{RS61}. If the region $\O$ contains
an open cone, it is well known to follow from conditions (A) through (C)
(cf., e.g., the Appendix in \cite{Buc74}). The Reeh-Schlieder property
holds as soon as one has {\em weak additivity} (\cite{Bor65}, cf. also Thm. 7.3.37 in 
\cite{BaW92}): if $\O$ is any bounded open region, then
$$\left(\bigcup_{a\in\Rd}\dA(\O+a)\right)''=\aloc''.$$
Conversely, it is well known that weak additivity can be derived from the 
Reeh-Schlieder property as well, provided Assumptions (A) through (C) hold
(cf., e.g., Lemma 2.6 in \cite{TW97a}). For the reader's convenience
we include a proof of this fact in the appendix.

Some special classes of spacetime regions will be used below. The first one
is the class $\K$ of {\em double cones}, i.e., all regions
of the form $(a+V_+)\cap(b-V_+)$, $a,b\in\Rd$. 
The class $\W$ of {\em wedges} consists of the region
$W_1:=\{x\in\Rd:\,x_1>|x_0|\}$ and its images under Poincar\'e
transforms. 

If $M$ is a region in $\Rd$, one denotes by $M^c$ the
{\em causal complement} or {\em spacelike complement}, which is the
region consisting of all points that are spacelike with respect to all
points of $M$.  The spacelike complement of the spacelike complement
$(M^c)^c=:M^{cc}\supset M$ is called the {\em causal completion} of
$M$, and $M$ is called {\em causally complete} if $M=M^{cc}$. It is
convenient to denote the interior of $M^c$ by $M'$.

$\K$ and $\W$ are subclasses of the class $\C$ of convex, causally
complete and open proper subsets of $\Rd$. 
The wedges in $\W$ are maximal elements of $\C$
in the sense that for every wedge $W\in\W$, every element $R\in\C$ 
with $R\supset W$ is
a wedge. Every element $R$ of $\C$ is an intersection of wedges
(cf. \cite{TW97}, Thm. 3.2)\footnote{For the proof of
this statement it is essential 
that $\C$ consists of open regions (the statement also implies to regions 
with a nonempty interior). As a 
counterexample, consider the lightlike half plane
$R:=\{x\in\Rd:\,x_1=x_0,\,x_2>0,\dots x_s>0\}$. One checks that
this region is causally complete and convex, while it is not an
intersection of wedges.}. The class of
all wedges that contain a region $R$ will be denoted by
$\W_R$. 

In general, the causal complement of a region in $\C$ is not
convex. If $R\in\C$, then $R'$
is a union of wedges (\cite{TW97}, Thm. 3.2), and
$\W^{R'}$ will denote the class of all wedges that are
subsets of $R'$. 
If $\O$ is an open convex region and if $P$ is a convex region that is spacelike
separated from $\O$, there is a wedge $X\in\W$
such that $\O\subset X$ and $P\subset W^c$ (cf. \cite{TW97}, Prop. 3.1).  

$\B$ will denote the bounded elements of the class $\C$. Clearly, the
double cones are in $\B$. Every element of $\B$ is contained in some
double cone, and it is precisely the intersection of all such double
cones (\cite{TW97}, Prop. 3.8). The class of all double cones which
contain a region $\O$ will be denoted by 
$\K_\O$, and the class of all double cones contained in an
arbitrary region $R$ will be called $\K^R$.

In Section \ref{nonempty-intersection}, two more technical assumptions will occur:
\begin{quote}
{\bf (E) Wedge duality.} For all $W\in\W$, one has $\dA(W')'=\dA(W)''$.

{\bf (F) Wedge additivity.}  For each
wedge $W\in\W$ and each double cone $\O\in\K$ with $W\subset W+\O$ one
has
$$\dA(W)''\subset\left(\bigcup_{a\in W}\dA(a+\O')'\right)''.$$
\end{quote}
All nets arising from finite-component Wightman fields satisfy wedge
duality \cite{BW75,BW76}. One checks that wedge duality implies the
condition of essential duality known from the analysis of
superselection sectors, since for any two spacelike separated regions
in $\B$, one can find a wedge which contains one of the two, whereas
its spacelike complement contains the other one
(see above, and cf. Lemma \ref{BoY} below). The algebras $\dA(\O')'$
will also occur below for other regions $\O$, e.g., for
double cones. By locality, $\dA(\O')'\supset
\dA(\O)$, and with the above remarks, it is easy to show that as
soon as wedge duality holds, one obtains that
for spacelike separated double cones $\O$ and $P$,
the elements of the algebras $\dA(\O')'$ and $\dA(P')'$ commute,
a property which is called {\em essential duality} and which is used
in the theory of localized superselection sectors (cf. \cite{Haa92}
and references given there).

Condition (F) strengthens Condition (D) slightly, but it is a standard
property of all Wightman fields as well. It has been used
extensively by Thomas and Wichmann 
in \cite{TW97a}. The authors have obtained results in the spirit of
Theorem \ref{Landau} and Proposition \ref{wed add suf} below, but their
results do not imply ours.

Occasionally, terminology borrowed from PDEs and General Relativity
will be used (timelike curves, Cauchy surfaces, etc.). These notions
will not be defined in detail, but will be used as in \cite{HE80}.

\subsection{Commutator functions and wave equation techniques}
\label{wave equation techniques}

It is a classical result of the Wightman approach to quantum field
theory that one can reconstruct a Wightman field from its vacuum
expectation values \cite{StW64,Jos65}. 
The following lemma shows how one can reconstruct commutation
relations of a net of observables from the behaviour of its vacuum
expectation values. Since these have some convenient properties, this
will facilitate the subsequent investigations. $\dA$ will be a local net
of local observables satisfying the above Assumptions (A) through (C).

\subsubsection{Lemma}\label{commutators}
\begin{quote}{\it
For an arbitrary double cone $\O\in\K$, let $A$ be an element of
$\dA(\O')'$. }

\begin{quote}{\it
(i) If a region $R\subset\Rd$ contains some
  open cone and has the property that
$\left<\gO,AB\gO\right>=\left<\gO,BA\gO\right>$ for all $B\in\dA(R)$,
then $A\in\dA(R)'$.

(ii) Assume that $\dA$ has the Reeh-Schlieder property, and 
suppose there is a double cone $P\in\K$ with the property that
$\left<\gO,AB\gO\right>=\left<\gO,BA\gO\right>$ for all $B\in\dA(P)$.
If there is a double cone $Q\subset P$ with the property that 
$A\in\dA(Q)'$, then $A\in\dA(P)'$.

(iii) Assume that $\dA$ exhibits the Reeh-Schlieder property, and
suppose there is a double cone $P\in\K$ with the property that
$\left<\gO,AB\gO\right>=\left<\gO,BA\gO\right>$ for all
$B\in\bigcup_{a\in\Rd}\dA(P+a)$. Then $A$ is a multiple of the
identity.}
\end{quote}
\end{quote}
{\bf Proof.} (i): If $S$ is an open cone
contained in $R$, there is a translation $a\in\Rd$ 
such that $S+a\subset R\cap\O'$.
Choose $C$ and $D$ in $\dA(S+a)$ and $B$ in $\dA(R)$. 
Since $A\in\dA(\O')'$, the operators $A$ and $C^*$ commute:
$$\langle C\gO,ABD\gO\rangle=\langle\gO,C^*A\,BD\gO\rangle=
\langle\gO,AC^*BD\gO\rangle.$$
Since $C^*BD$ is in $\dA(R)$, the assumption implies
$$\langle\gO,AC^*BD\gO\rangle=\langle\gO,C^*BDA\gO\rangle,$$
and since $D$ and $A$, in turn, commute because of $A\in\dA(\O')'$, 
one concludes
$$\langle C\gO,ABD\gO\rangle=\langle\gO,C^*BDA\gO\rangle=
\langle C\gO,BAD\gO\rangle.$$
But since $C$ and $D$ are arbitrary elements of $\dA(S+a)$, and since
$\gO$ is cyclic with respect to this algebra,
it follows that $AB=BA$; since
$B\in\dA(R)$ was arbitrary, one obtains $A\in\dA(R)'$, which is (i). 

(ii) Choose $C$ and $D$ in $\dA(Q)$ and $B$ in $\dA(P)$. 
Since $A$ has been assumed to be in $\dA(Q)'$, it commutes with $C^*$, so
$$\langle C\gO,ABD\gO\rangle=\langle\gO,C^*A\,BD\gO\rangle=
\langle\gO,AC^*BD\gO\rangle.$$
Since $C^*BD$ is in $\dA(P)$, the assumption implies
$$\langle\gO,AC^*BD\gO\rangle=\langle\gO,C^*BDA\gO\rangle,$$
and since $D$ and $A$ commute by the assumption that $A\in\dA(Q)'$, 
one concludes
$$\langle C\gO,ABD\gO\rangle=\langle\gO,C^*BDA\gO\rangle=
\langle C\gO,BAD\gO\rangle.$$
But since $C$ and $D$ are arbitrary elements of $\dA(Q)$, and since
by the Reeh-Schlieder property, $\gO$ is cyclic with
respect to this algebra, 
it follows that $AB=BA$; since
$B\in\dA(P)$ was arbitrary, one obtains $A\in\dA(P)'$, which is (ii). 

(iii) There is a translation $a\in\Rd$ 
such that $P+a\subset\O'$, so that $A\in\dA(\O')'\subset\dA(P+a)'$. 
Now choose a $b\in\Rd$ such that $P+b$ intersects $P+a$, and let $Q$
be a double cone contained in $(P+b)\cap(P+a)$. Isotony implies that
$A\in\dA(Q)'$. Since by assumption, 
$\langle\gO,AB\gO\rangle=\langle\gO,BA\gO\rangle$ for all
$B\in\dA(P+b)$, (ii) implies that $A\in\dA(P+b)'$. Now one can iterate
this procedure: choose an arbitrary $c\in\Rd$ such that $(P+c)\cap(P+b)$
is nonempty, choose a new double cone $Q$ in this intersection, and
conclude from (ii) that $A\in\dA(P+c)'$.
Note that only the 
double cone $P+a$ chosen in the first step 
needs to be spacelike separated from $\O$, since
each step uses the result of the preceding one, so one finds that for
{\em every} $a\in\Rd$, one proves that $A\in\dA(P+a)'$ with a finite
number of steps. 
The statement now follows from weak additivity, which follows from
the Reeh-Schlieder property (see above), and from irreducibility.

\Halmos

\bigskip\bigskip\noindent%
Given any two local observables $A,B\in\aloc$, the 
commutator function $f_{A,B}$ will henceforth be defined by
$$\Rd\ni x\mapsto\langle\gO,[A,U(x)BU(-x)]\gO\rangle=:f_{A,B}(x).$$
Due to Lemma \ref{commutators}, the analysis of the support of
this function yields
information on the structure of the net. Crucial for this analysis
is the fact that $f_{A,B}$ is a boundary value of a solution of 
the wave equation, and a well-known lemma due to Asgeirsson
concerning such solutions (cf., \cite{Ara63}, Sect. 4.4.D in \cite{BLOT90}, 
or \cite{CH II}) immediately implies the following lemma,
which, for this reason, will be referred to as Asgeirsson's
Lemma. Another important consequence of the \lq wave nature\rq\, of the
function $f_{A,B}$ is a theorem due to Jost, Lehmann and Dyson \cite{JL57,Dys58} 
which will also be recalled for the reader's convenience. 

\subsubsection{Lemma (Asgeirsson)}\label{Asg}
\begin{quote}{\it
If the commutator function $f_{A,B}$ and all its partial derivatives
are zero along a 
timelike curve segment $\gg$, $f_{A,B}$ vanishes in the entire double cone
$\gg^{cc}$.}
\end{quote}
{\bf Proof.} The Fourier transform of the 
operator valued function $\Rd\ni x\mapsto U(x)$ 
is the spectral measure of the four-momentum operator. It follows
that the Fourier transform $\Hat{f}_{A,B}$ of the function
$f_{A,B}$ is a finite (not necessarily positive) 
measure, and by the spectrum condition, one has  
$\supp\,\hat f_{A,B}\subset\overline{V}$. It follows that the function
$$F(x,\sigma):=(2\pi)^{-\frac{1+s}{2}}\int \cos(\sigma
\sqrt{k^2})\,e^{ikx}\,d\hat f_{A,B}(k)$$
is a continuous function with $F(x,0)=f_{A,B}(x)$ for all
$x\in\Rd$. This $F$ is a solution of the 1+(s+1)-dimensional wave
equation. This implies the statement by Asgeirsson's result for solutions of 
the wave equation, see the references quoted above.

\Halmos

\bigskip\bigskip\noindent%
Evidently, the assumption of the lemma is satisfied as soon as $f_{A,B}$ vanishes in 
some open neighbourhood of $\gamma$.
In the proof of Theorem \ref{JLD} below, however, the function $F$ defined in
the proof is analysed, and the information one has about $f_{A,B}$
from locality merely implies that $F$ vanishes in a null set of 
$\reals^{1+(s+1)}$. In this case one 
makes use of the fact that $F$ has been
constructed in such a way that all its partial derivatives, including
the one in the $\gs$-direction, are zero at all points of this null set; 
one may then use the above lemma to show that the region
where $F$ vanishes also extends into the $\gs$-direction. 

\subsubsection{Definition}
\begin{quote}
Let $R$ be a region in Minkowski space.

(i) $R$ will be called {\bf Asgeirsson complete} if for every 
timelike curve segment $\gamma\subset R$, the double cone $\gamma^{cc}$ 
is a subset of $R$ as well. The smallest
Asgeirsson complete extension of $R$ will be called the {\bf Asgeirsson
hull} of $R$. 

(ii) $R$ will be called {\bf timelike convex} if it
contains as subsets all
double cones with tips in $R$, i.e., if
$(R+V_+)\cap(R-V_+)\subset R$. 

(iii) $R$ will be
called a {\bf Jost-Lehmann-Dyson region} if it is timelike convex
and if every inexdentible timelike curve in $\Rd$ intersects $R\cup R^c$.
\end{quote}
Timelike convex regions contain all
timelike path segments connecting two points in the region, so the
terminology is in harmony with other notions of convexity. In
\cite{Kuc98} the term \lq double cone complete\rq\, was used instead
of \lq timelike convex\rq, but the latter
term was also used in \cite{TW97} (Par. IV) and
will be used in what follows to facilitate reading.
The following lemma collects some relations between these notions
most of which will be used below.

\subsubsection{Lemma}\label{region}

\begin{quote}{\it
(i) Every causally complete region is timelike convex.

(ii) Every timelike convex region is Asgeirsson complete.

(iii) Every timelike convex and bounded open region 
is a Jost-Lehmann-Dyson region.

(iv) The causal complement of a Jost-Lehmann-Dyson
region is a Jost-Lehmann-Dyson region. 

(v) Let $R$ and $S$ be timelike convex regions, and assume that there
exists a Cauchy surface $T$ which
is a subset of both $R$ and $S$. Then 
the region $R\cup S$ is timelike convex (and, like $R$ and $S$, 
trivially, a Jost-Lehmann-Dyson region). 

(vi) Let $(R_\rho)_{\rho>0}$ be an increasing family of Jost-Lehmann-Dyson
regions. Then $R:=\bigcup_{\rho}R_\rho$ is a Jost-Lehmann-Dyson region.}
\end{quote}
Before proving the lemma, we give some counterexamples to strengthened statements or
converse implications.
An example of a timelike convex region (and Jost-Lehmann-Dyson region)
that is not causally complete (cf. (i))
is the time slice region $\{x\in\Rd:\,0\leq x_0\leq1\}$. An example of an 
Asgeirsson complete region that is not timelike convex (cf. (ii)) is the
union of two disjoint double cones at a timelike distance; this 
shows that the classes of timelike convex regions and of Jost-Lehmann-Dyson
regions, respectively, are not stable under taking unions,
so Statement (v) is far from tautological.
The same holds for the class of Asgeirsson complete regions: consider the 
regions $R_+:=\{x\in\Rd:\,\rho x_1<x_0<\rho x_1+1\}$ and $R_-:=\{x\in\Rd\,
-\rho x_1<x_0<1-\rho x_1\}$ for some $\rho$ with $0<\rho\leq 1$. 
One easily checks that both regions are Asgeirsson
complete, while their union is not: its Asgeirsson hull is $\Rd$.
If $\rho<1$, the two regions are even Jost-Lehmann-Dyson regions, 
while their union evidently is not (cf. (v) and (vi)).

An example of a timelike convex region which is neither causally
complete nor a Jost-Lehmann-Dyson region (cf. (iii))is the region 
$$R:=\{x\in\Rd:\,1<x^2<2, x_0>0\},$$
since there are timelike curves which do not intersect $R$, e.g.,
the curve $\reals\ni t\mapsto(\sinh t,\cosh t,0,\dots,0)$. 

\bigskip\bigskip\noindent%
{\bf Proof of Lemma \ref{region}.} 
(i) Let $R$ be a causally complete region, and pick two points
$x\in R$ and $y\in R\cap (x+V_+)$. The causal 
completion $\{x,y\}^{cc}$ of the set $\{x,y\}$ is the closure of the
double cone $(x+V_+)\cap(y-V_+)$, 
and since $\{x,y\}\subset R$ implies $\{x,y\}^{cc}
\subset R^{cc}=R$, this immediately implies (i).

Statement (ii) immediately follows from the definition. 

(iii) Let $R$ be timelike convex, bounded and open.
Since $R$ is open, a point $x\in\Rd$ is 
not contained in the spacelike complement $R^c$
if and only if it is timelike with respect to some point
of $R$, i.e., $\Rd\backslash R^c=R+V$, where $V$ is the open light cone. Now
let $\gamma$ be an inextendible timelike curve that does not intersect $R\cup R^c$. 
Since $\gamma$ does 
not intersect $R^c$, it has to stay within the region $R+V$.
But since $\gamma$ is an inextendible timelike curve, while $R$
is bounded, $\gamma$ cannot stay in 
the future $R+V_+$ or the past $R-V_+$ of $R$, i.e., it 
has to pass from $R-V_+$ to $R+V_+$. Since both these regions are
open, while $\gamma$ is continuous, it follows that it has to 
hit the region $(R+V_+)\cap(R-V_+)$. But this region coincides with $R$
since $R$ is timelike convex and open, so 
$\gamma$ hits $R$, which is a contradiction and proves (iii).

(iv) The causal complement of any region is causally complete, by (i), 
this enhances timelike convexity of $R^c$. 
The condition that $R\cup R^c$ 
is intersected by every inextendible timelike curve implies that 
$R^c\cup R^{cc}$ ($=R^c\cup R$)
is intersected by every such curve. This proves (iv).

(v) Let $x$ and $y$ be points in $R\cup S$ that are timelike with respect 
to each other. Since $\Rd$ is timelike convex, one finds an
inextendible timelike curve $\gamma$ hitting both
$x$ and $y$. Let $z$ be the unique point where $\gamma$ hits $T$.
Since $R$ and $S$ are timelike convex, and since $z\in T\subset R\cap S$,
the closed double cones with the tips $z$ and $x$ and the 
tips $z$ and $y$, respectively, are subsets in $R\cup S$.
If with respect to the time ordering along $\gamma$, $z$ is earlier or later
than both $x$ and $y$, it follows that the double cone with tips 
$x$ and $y$ is contained in $R\cup S$ as well. If $z$ is between $x$ and $y$,
then, as before, we can conclude that the segments of $\gamma$ 
between $z$ and $x$
and between $z$ and $y$ is a subset of $R\cup S$, and since 
$z\in T\subset R\cap S$, it follows that all of 
the segment of $\gamma$ joining $x$ to $y$ is a subset of $R\cup S$.
Since $\gamma$
can be {\em any} inextendible timelike curve hitting $x$ and $y$, one obtains 
that all timelike curve segments joining $x$ and $y$ are contained in
$R\cup S$, so the double cone with tips $x$ and $y$
is contained in $R\cup S$, which completes the proof of (v).

(vi) Let $x$ and $y$ be two points in $R$ at a timelike distance. There are 
a $\rho_x>0$ and a $\rho_y>0$ such that $x\in R_{\rho_x}$ and
$y\in R_{\rho_y}$, so it follows that both $x$ and $y$ are elements of 
$R_{max\{\rho_x,\rho_y\}}$. Since this region is timelike convex,
it follows that the double cone with tips $x$ and $y$ is in $R$, proving
that $R$ is timelike convex.

It remains to be shown that every inextendible timelike curve intersects 
$R\cup R'$. Let $\gamma$ be an inextendible timelike curve that does not
intersect $R$. Since all $R_\rho$ are Jost-Lehmann-Dyson regions, it
follows that $\gamma$ has to intersect every $R_\rho'$, so it has
to intersect the region $\bigcap_{\rho>0}R_\rho'=R'$. This completes
the proof.

\Halmos

\bigskip\bigskip\noindent%
The above statements and proofs can be extended in a straightforward
manner to the spacetime one obtains by endowing the cyclinder
$$Z_\rho:=\{x=(x_0,\vec x)\in\Rd:\,\|\vec x\|=\rho\}$$ with the
spacetime structure it inherits from $\Rd$, provided $s\geq2$. For
$s=1$ this spacetime fails to be timelike convex, and the proof of
part (v) does no longer work.  For further results of the above kind,
see \cite{TW97}. The useful property of Jost-Lehmann-Dyson regions
(which is the reason to call them so) is established by the following
theorem.

\subsubsection{Theorem (Jost, Lehmann, Dyson)}\label{JLD}
\begin{quote}{\it
Let $A$ and $B$ be local observables, and assume that the commutator function 
$f_{A,B}$ vanishes in a Jost-Lehmann-Dyson region $R$.
Then the support of $f_{A,B}$ is contained not only in the complement of
$R$, but even in the (in general, smaller) union of all {\bf admissible
mass hyperboloids of $R$}, i.e., the mass hyperboloids 
$$H_{a,\gs}:=\{x\in\Rd:\,(x-a)^2=\gs^2\},\qquad a\in\Rd,\,\gs\in\reals,$$ 
which do not intersect $R$.}
\end{quote}
{\bf Sketch of proof.}
Define $F$ as in the proof of Lemma \ref{Asg}. Since $F$
is a solution of the wave equation, it is well-known that 
for every Cauchy surface $\zeta$ in $\reals^{1+(s+1)}$, there exists
a distribution $F_\zeta$ with support in
$\zeta$ such that $F=F_\zeta*D_{1+(s+1)}$, where $D_{1+(s+1)}$ denotes
a fundamental solution of the 1+(s+1)-dimensional wave equation
(see, e.g., \cite{BLOT90}, pp. 175-184). The support of $D_{1+(s+1)}$
is contained in the closed light cone $\overline{\widehat V}$ of
$\reals^{1+(s+1)}$\,\,\footnote{This notation is consistent since 
  $\widehat V$ is, indeed, the 1+(s+1)-dimensional Asgeirsson hull of
  $V$. Note that $\widehat{\overline{V}}\not=\overline{\widehat V}$.}. 
Since $R$ is a Jost-Lehmann-Dyson region in $\Rd$,
its 1+(s+1)-dimensional Asgeirsson hull $\widehat R$ is easily seen to be  
a Jost-Lehmann-Dyson region in $\reals^{1+(s+1)}$.
Provided this region is \lq well-behaved\rq, there is a
Cauchy surface $\zeta$ in $\widehat R\cup\widehat R^c$. 
This Cauchy surface has the property that for every point $z\in\zeta$, either
both the forward and the backward part of $\overline{\widehat{V}}+z$ 
or neither of them intersects $R$. The former case occurs if and only if 
$z\in\zeta\cap\widehat R$.
The latter case occurs if and only if $z\in\zeta\cap\widehat R^c$, 
the Asgeirsson hull $\widehat R$ of $R$ and the spacelike complement 
being taken in the spacetime $\reals^{1+(s+1)}$. 
But since all partial derivatives of $F$ can be checked to
vanish in all points in $R$, one obtains from Lemma \ref{Asg} that
$F$ vanishes in $\widehat R$, the support of $F_\zeta$ contains only
points of the second kind, i.e., 
it is contained in $\widehat R^c\cap\zeta$. This implies
that the support of $F$ is contained in $(\widehat
R\cap\zeta)+\overline{\widehat V}$. 

Since $f_{A,B}$ is a boundary value of $F$ and 
since the intersection of $\overline{\widehat{V}}+c$ with $\Rd$ is the
convex hull of a shifted mass hyperboloid, the support of the 
boundary value $f_{A,B}$ of the function $F$ is contained in the union
of admissible mass hyperboloids, as stated.

\Halmos

\bigskip\bigskip\noindent%
We conclude this section with another lemma to be used below that concerns
the geometry of Minkowski space.

\subsubsection{Lemma}\label{ominusp}
\begin{quote}{\it
Let $P\in\K$ be a double cone.

(i) If $\O$ is a double cone, so is $(\O+P)^{cc}$.

(ii) If $W$ is a wedge, so is $(W+P)^{cc}$.}
\end{quote}
{\bf Proof.} Denote by $a_{\O}$ and $a_P$ the lower tips,
and by $b_\O$ and $b_P$ the upper tips of $\O$ and $P$, respectively.
Let $x=a_\O+\xi$ and $y=a_P+\eta$ be points in $\O$ and $P$, respectively.
Then $x+y=a_\O+a_P+\xi+\eta$, and since $\xi$ and $\eta$ are elements of
$V_+$, so is $\xi+\eta$, so $x+y$ is contained in $a_\O+a_P+V_+$. In the 
same way one proves that $x+y\in b_\O+b_P-V_+$, so one has
$$\O+P\subset (a_\O+a_P+V_+)\cap(b_\O+b_P-V_+).$$
Since the right hand side is a double cone and, hence, causally complete,
it follows that $(\O+P)^{cc}$ is a subset of this double cone as well.
On the other hand it is straightforward to check that the tips $a_\O+a_P$
and $b_\O+b_P$ of this double cone and the straight line joining them 
are contained in $\O+P$, whence the converse 
inclusion follows as well, so the proof of (i) is complete.

The region $W+P$ is a union of wedges that are images of $W$
under translations. Consequently, $(W+P)^c$ is the intersection of
the corresponding translates of $W^c$ under translations. But this
intersection is the closure of a wedge, so it follows that the causal
complement of this region, $(W+P)^{cc}$, is a wedge. This proves (ii).

\Halmos

\section{Results}
\label{results}

By definition, a local net associates algebras with regions. 
In the sequel it will be discussed how to associate a localization
region with a given algebra and even with a single local observable. 
The analysis is based on a theorem
due to Landau \cite{Lan69}. In order to localize single observables, a
new generalization of Landau's theorem will be used. It will be stated
and proved below.

This section is structured as follows: 
in Section \ref{empty-intersection}, the theorem of Landau
and its consequences for the localization of algebras will be
discussed, and the mentioned generalization will be proved. 
This generalization
will be the basis for the analysis of localization regions for single
local observables, which is presented in Section \ref{nonempty-intersection}.

In what follows, Assumptions (A) through (D) will be made without
further mentioning, and it will assumed in addition that $s\geq2$;
Landau's theorem and all generalizations discussed below heavily depend on
this assumption, and so do the consequences to be discussed later on.

\subsection{Landau's theorem and the empty-intersection theorem}

\label{empty-intersection}

Using the wave equation techniques discussed in the preceding section, 
Landau \cite{Lan69} proved the following:

\subsubsection{Theorem (Landau)}\label{Landau0}
\begin{quote}
{\it If the closures of two double cones $\O$ and $P$ are disjoint,
  then}
$$\dA(\O')'\cap\dA(P')'=\complex\,\id_\H.$$
\end{quote}
\bigskip\noindent%
This already implies that for an $\O$ satisfying the assumptions of
the corollary, the region
$$L(\dA(\O')'):=\bigcup\{P\in\K:\,\dA(P)\subset\dA(\O')'\},$$ 
which will be called the {\bf localization region} of the algebra
$\dA(\O')'$, coincides with $\O$ (cf. \cite{Ban87}):

\subsubsection{Corollary}\label{Landau 2}
\begin{quote}{\it
Let $\O\subset\Rd$ be a bounded, causally complete and convex open region.

(i) For every open region $M\subset\Rd$, one has
$\dA(M)\subset\dA(\O')'$ if and only
if $M\subset\O$.

(ii) $L(\dA(\O))=\O$.
}
\end{quote}
{\bf Proof.} By isotony and locality, 
the condition in statement (i) is sufficient. To prove that it
is necessary, assume $M\not\subset\O$. 
Then, since $\K$ is a 
topological base and since the region
$M\backslash\overline{\O}$ has a nonempty interior, 
$M\backslash\overline{\O}$ contains a double cone
$P\in\K$ whose closure is disjoint from $\overline{\O}$. Since 
$\overline{\O}$ is an intersection of closures of 
wedges, it follows from this
that a wedge $W$ can be found whose closure is disjoint from
$\overline{P}$ and contains $\overline{\O}$. Since 
$\overline{P}$ is compact, the distance between $\overline{P}$
and $\overline{W}$ is $>0$, so eventually shifting it a little bit,
one can choose $W$ in such a way that $\overline{W}$
is a subset not only of $\overline{W}$, but also of $W$ itself.

By Proposition 3.8 (b) in \cite{TW97}, on can now conclude that 
there is a double cone $Q$ with 
$Q\subset W$ and $Q\supset\O$ (note that $\O$
itself does
not need to be a double cone). Landau's theorem now implies
that $\dA(P)\cap\dA(Q')'=\complex\,\id_{\H}$. It follows from the
Reeh-Schlieder property that $\dA(P)\not\subset\complex\,\id_{\H}$,
so $\dA(P)\not\subset\dA(Q')'$. Since 
$\dA(P)\subset\dA(M)$ follows from isotony, $\dA(M)$
cannot be a subset of $\dA(Q')'$, and since $\O\subset Q$, it cannot be a
subset of $\dA(\O')'$. This proves (i) and, trivially, implies (ii).

\Halmos  

\bigskip\bigskip\noindent%
The proof of Corollary \ref{Landau 2} can be made
shorter as soon as one knows that Landau's theorem still
works if one of the two double cones is replaced by a wedge.
That this, indeed, is possible, has been shown in the context of the
proof of the \pct-part of the first uniqueness theorem for modular
symmetries (Theorem 2.1 in \cite{Kuc97}). 

\subsubsection{Theorem}
\begin{quote}{\it
If the closures of a double cone $\O$ and a wedge $W$ are disjoint,
then}
$$\dA(\O')'\cap\dA(W')'=\complex\,\id_\H.$$
\end{quote}
\bigskip\noindent%
Using this generalized
version of Landau's theorem, one concludes that in Lemma
\ref{Landau 2}, the assumption that $\O$ is bounded may be omitted:

\subsubsection{Corollary}\label{Landau 2a}
\begin{quote}{\it
Let $R\subset\Rd$ be a causally complete convex open region.

(i) For every open region $M\subset\Rd$, one has
$\dA(M)\subset\dA(R')'$ if and only
if} $M\subset R$.

(ii) $L(\dA(R))=R$.
\end{quote}
{\bf Proof.} By isotony and locality, 
the condition is sufficient. To prove that it
is necessary, assume $M\not\subset R$. 
Then, since $\K$ is a topological base and since the region
$M\backslash\overline{R}$ has a nonempty interior, $M\backslash 
\overline{R}$ 
contains a double cone
$\O\in\K$ whose closure is disjoint from $\overline{R}$. As in the
proof of Corollary \ref{Landau 2}, it follows
that a wedge $W$ can be found whose closure is disjoint from
$\overline{\O}$ and whose interior contains $\overline{R}$. 
Landau's theorem now implies
that $\dA(\O)\cap\dA(W')'=\complex\,\id_{\H}$. It follows from the
Reeh-Schlieder property that $\dA(\O)\not\subset\complex\,\id_{\H}$,
so $\dA(\O)\not\subset\dA(W')'$. Since 
$\dA(\O)\subset\dA(M)$ follows from isotony, $\dA(M)$
cannot be a subset of $\dA(W')'$, and since $R\subset W$, it cannot be a
subset of $\dA(R')'$, proving both statements.

\Halmos  

\bigskip\bigskip\noindent%
In order to investigate the localization behaviour of a single local
observable, a further generalization of Landau's
theorem will be used. It is the main result of this section. It 
is a generalization of Theorem 2.1 in \cite{Kuc97}. $\aloc^d$ will
denote the algebra of local observables of the dual net $\dA^d$.

\subsubsection{Theorem (empty-intersection theorem)}\label{Landau}
\begin{quote}{\it
Let $(W_\nu)_{1\leq\nu\leq n}$ be a family of $n$ wedges in $\W$.
If $\bigcap_{\nu}\overline{W}_\nu=\emptyset$, then} 
$$\aloc\cap\bigcap_{\nu}\dA(W_\nu')'=\complex\,\id_{\H}.$$
\end{quote}
{\bf Proof.} Choose an $A\in\aloc\cap\bigcap_\nu\dA(W_\nu')'$, and
let $\O$ be a double cone with $A\in\dA(\O)$ (or $A\in\dA(\O')'$).

Since the wedges $\overline{W}_{\nu}$ have empty common intersection,
so do the compact regions $\overline{\O}\cap\overline{W}_{\nu}$. But
if a finite family of compact regions have empty common intersection,
there is an $\eps>0$ such that the family of the $\eps$-neighbourhoods
of the regions still have empty common intersection. The proof of this is
an elementary induction proof: any two disjoint compact regions have a
positive distance, which implies the statement for two regions. Now assume
the statement to hold for any family of $n$ compact sets, and
let $C_1,\dots,C_{n+1}$ be a family of n+1 regions. If $n$ of these
regions already have empty common intersection, there is nothing more
to prove. So consider the case that the set $\Gamma:=
\bigcap_{\nu=1}^n C_\nu$ is nonempty. This region is compact and, as
shown, has a finite distance $\delta$ from $C_{n+1}$, so the $\delta/3$
neighbourhoods of the two regions still have a finite distance.
But the $\delta/3$-neighbourhood of $\Gamma$ is the intersection 
of the $\delta/3$-neighbourhoods of $C_1,\dots,C_n$, so the statement
follows for $\eps=\delta/3$.

It follows from this that
there is a double cone $P$ which is so small that the wedge
$\tilde W_\nu:=(W_\nu-P)''=(W_\nu-P)^{cc}$, $\nu\leq n$, and the
double cone $\tilde\O:=(\O-P)''=(\O-P)^{cc}$ (cf. Lemma \ref{ominusp} above to 
see that these regions are a wedge and a double cone, respectively)
still have empty 
common intersection. Choose any $B\in\dA(P)$. By locality,
the commutator function $f_{A,B}$ vanishes
in the region $R:=\tilde\O'\cup\bigcup_\nu\tilde W_\nu'$. 

There is no admissible mass hyperboloid for this region. To see
this, note that if a (shifted) mass hyperboloid is disjoint from 
a union of wedges, so is the unique shift
$x+\overline{V}$, $x\in\Rd$, of the closure of the full light cone
which contains the hyperboloid.
Now choose $x\in\Rd$ such that $x+\overline{V}$ is disjoint from
all $\tilde W_\nu'$, $\nu\leq n$, and from $\tilde\O'$, which
is a union of wedges, too. This is equivalent to
$\{x\}'\supset\tilde\O'\cup\bigcup_{\nu}\tilde W'_\nu$, i.e.,
$$x\in\tilde\O''\cap\bigcap_{\nu}\tilde W''_\nu
=\tilde\O\cap\bigcap_{\nu}\tilde W_\nu=\emptyset.$$
Hence there is no admissible mass hyperboloid for $R$.

If $R$ is a Jost-Lehmann-Dyson region, it follows from 
Theorem \ref{JLD} that $f_{A,B}(x)$ vanishes for all $x\in\Rd$ and all
$B\in\dA(P)$,
so using part (iii) of Lemma \ref{commutators}, one concludes that 
$A\in\complex\,\id_\H$, and the proof is complete.

But since $R$ does not need to be a Jost-Lehmann-Dyson region,
Asgeirsson's lemma will be used to show that the function $f_{A,B}$ 
vanishes in a larger region $N\supset R$ which is a Jost-Lehmann-Dyson 
region. Since there is no admissible hyperboloid
for $R$, there is, a fortiori, no admissible
hyperboloid for $N$, so the proof will be complete as soon as
$N$ has been shown to exhibit the stated properties.

To this end, choose coordinates such that $\tilde\O$ is the double cone
$$(-\rho_0 e_0+V_+)\cap(\rho_0 e_0-V_+),$$
where $e_0$ denotes the unit vector in the 0-direction, and
$\rho_0>0$ is the radius of the double cone $\tilde\O$.
Let $Z_\rho=\{x=(x_0,\vec x)\in\Rd:\|{\vec x}\|=\rho\}$ be the
boundary of the cylinder of radius $\rho$ around the time axis in
$\Rd$, and define
\begin{eqnarray*}
R_{\rho,0}&:=&\tilde\O'\cap Z_\rho,\\
R_{\rho,\nu}&:=&\tilde W_\nu'\cap Z_\rho,\qquad\nu\leq n;
\end{eqnarray*}
All these regions are bounded subsets of $\Rd$.
Due to our choice of coordinates, the region $R_{\rho,0}$
is a strip: 
$$R_{\rho,0}=\{x\in Z_\rho:\,|x_0|\leq \rho-\rho_0\}$$
(which is empty if $\rho<\rho_0$). 

For $1\leq\nu\leq n$, the
wedge $\tilde W'_\nu$ is timelike convex in $\Rd$, so the region
$R_{\rho,\nu}$ is timelike convex
with respect to the inherited spacetime structure of $Z_\rho$. 
We now show that there is a $\rho_\nu>0$ such that 
the union $R_{\rho,0}\cup R_{\rho,\nu}$
is timelike convex as well for all $\rho>\rho_\nu$. To this end, 
let $C$ be a spacelike hypersurface in $\tilde W_\nu\cup\tilde W_\nu^c$. 
As a spacelike surface, it is a subset of $\tilde\O'$ up to a
compact set. For $\rho$ so large that this compact set is 
enclosed by $Z_\rho$ one finds that $C\cap Z_\rho$ is a subset of
$R_{\rho,0}$. Since $C\cap Z_\rho$ is a Cauchy surface in the spacetime $Z_\rho$, 
it follows that $R_{\rho,\nu}\cup C$ and $R_{\rho,0}$
are timelike convex regions in the spacetime $Z_\rho$
whose intersection contains a Cauchy surface,
so part (v) of Lemma \ref{region} implies that $R_{\rho,0}\cup R_{\rho,\nu}$
is timelike convex. This proves that $\rho_\nu$ with the stated properties 
exists for $1\leq\nu\leq n$.

Now choose $\rho>\hat\rho:=\max_\nu \rho_\nu$, 
and apply Lemma \ref{region} (v) another $n-1$ times to conclude that the region
$$R_\rho:=R\cap Z_\rho=\bigcup_{0\leq\nu\leq n}R_{\rho,\nu}$$ 
is timelike convex in $Z_\rho$.
Since the $\Rd$-Asgeirsson hull $\widehat R_\rho$
is open, bounded, and timelike convex, it is a 
Jost-Lehmann-Dyson region by Lemma \ref{region} (iii). 

On the other hand, the part of $\tilde\O'$ and 
$\tilde W_\nu'$, respectively, which is enclosed by $Z_\rho$
is a subset of the $\Rd$-Asgeirsson hull
$\widehat R_{\rho,\nu}$ of $R_{\rho,\nu}$. It follows that
$$R\subset N:=\bigcup_{\rho>\hat\rho}\bigcup_{\nu\leq n}
\widehat R_{\rho,\nu}
=\bigcup_{\rho>\hat\rho}\widehat R_\rho,$$
and by Asgeirsson's lemma, $f_{A,B}$ vanishes in $N$.
Since the Jost-Lehmann-Dyson region $\widehat R_\rho$ increases with $\rho$, 
it follows from Lemma \ref{region} (vi) 
that $N$ is a Jost-Lehmann-Dyson region.
This is what remained to be shown, so the proof is complete.

\Halmos 

\bigskip\bigskip\noindent%
Actually, the following, slightly stronger version has been established
by the preceding proof:

\subsubsection{Corollary}\label{empty2}
\begin{quote}{\it
Let $W_1,\dots W_n$ be wedges in $\W$, and let $\O\in\K$ be a double cone. 
If $\overline{\O}\cap\bigcap_{1\leq\nu\leq n}
\overline{W}_\nu=\emptyset$, then 
$$\dA(\O')'\cap\bigcap_\nu\dA(W_\nu')'=\complex\id.$$}
\end{quote}
\bigskip\bigskip\noindent%
After completing this article, it was brought to the author's
attention that in 1+3 dimensions, one can also use the results of
Thomas and Wichmann for the above proof. As the region
$\tilde\O'\cap\bigcap_\nu\tilde W_\nu'$ is a union of wedges, one can
apply Theorem 3.6 in \cite{TW3} to prove that the function $f_{A,B}$
vanishes in the causal closure of this region, which one can check to
be all of $\Rd$. Their proof, which uses a completely different line
of argument, has been written down for 1+3 dimensions only, and it is
not evident whether it also holds in other dimensions; verifying this
would require to check several hard proofs at the end of
\cite{TW97a}. On the other hand the Thomas-Wichmann analysis is more
general in other aspects, so the reader interested in the above
special problem may find the above argument more direct.

\subsection{The localization region of a single local observable and 
the nonempty-intersection theorem}
\label{nonempty-intersection}

Theorem \ref{Landau} prepares for the definition of a localization region for
local observables. As the following proposition shows, there are several 
\lq natural\rq\, ways how to define such a localization region, and it follows
from the empty-intersection theorem that all of them yield nonempty
localization regions.

\subsubsection{Proposition}\label{nonempty}
\begin{quote}{\it
Let $\X$ be any of the classes $\K$, $\B$, $\W$ and $\C$. For every
$A\in\aloc$ which is not a multiple of the identity, the {\bf
  localization regions}
\begin{eqnarray*}
{\bf L}^\X(A)&:=&\bigcap\{\overline{\O}:\,\O\in\X:\,A\in\dA(\O)''\}\\
L^\X(A)&:=&\bigcap\{\overline{\O}:\,\O\in\X:\,A\in\dA(\O')'\}
\end{eqnarray*}
are nonempty, causally complete, convex, and compact sets. 
Between them, one has the following equalities and inclusions:}
$$\begin{array}{ccccccc}
{\bf L}^\B(A)&=& {\bf L}^\K(A)&\supset&L^\K(A)&=&L^\B(A)\\
& &\cup& &\cup& &\\
{\bf L}^\C(A)&=&{\bf L}^\W(A)&\supset&L^\W(A)&=&L^\C(A)
\end{array}$$
\end{quote}
{\bf Proof.} We start with the proof of the equalities and inclusions.
The equalities immediately follow from the definitions,
since on the one hand, $\K\subset\B$ and $\W\subset\C$, while on the
other hand,
every region in $\B$ is an intersection of double cones in $\K$
and every region in $\C$ is an intersection of wedges in $\W$ (see
Section \ref{notation}). The inclusions in the upper
and the lower row of the diagram immediately follow from locality. The
inclusions in the two columns follow from the fact that every double
cone is an intersection of wedges and that, by isotony, an observable
contained in the algebra associated with a given double cone is 
contained in all algebras associated with wedges containing this
double cone. 

By these inclusions, it is sufficient to prove that $L^\W(A)$ is
nonempty if $A\notin\complex\,\id$. It already follows from Theorem
\ref{Landau} that the intersection of the closures of any
finite family of wedges
whose algebras contain $A$ is nonempty. But the family of all wedges
whose algebras contain $A$ is never finite. 

Since $A$ is a local observable,
there is a double cone $\O$ with $A\in\dA(\O)$, and it follows from
isotony, locality, and the above inclusions 
that $L^\W(A)\subset\overline{\O}$. 
But this implies that
$$L^\W(A)=\bigcap
\lbrace\overline{\O}\cap \overline{W}:\,W\in\W,\,A\in\dA(W')'\rbrace,$$
which is an intersection of subsets of the compact set
$\overline{\O}$. But if for a class of closed subsets of a compact
space, every finite subclass has a nonempty intersection, it follows
from the Heine-Borel property
that the whole class has a nonempty intersection. Now Corollary 
\ref{empty2} implies the statement.

\Halmos

\bigskip\bigskip\noindent%
In the sequel the maps $\aloc\ni A\mapsto {\bf L}^\X(A)$ and $\aloc\ni
A\mapsto L^\X(A)$ will be referred to as {\bf localization
  prescriptions}. 
Clearly, the localization prescriptions
${\bf L}^\K$ and $L^\K$ coincide if the net satisfies Haag duality, i.e.,
if $\dA(\O')'=\dA(\O)$ for all $\O\in\K$, and
the prescriptions ${\bf L}^\W$ and $L^\W$ coincide if the net satisfies
wedge duality. Furthermore, wedge duality also makes $L^\W$ coincide
with $L^\K$ by the following lemma (cf. also \cite{BoY94}, Lemma 4.1).

\subsubsection{Lemma}\label{BoY}
\begin{quote}
{\it 
Assume the net $\dA$ to satisfy wedge duality. For every
region $R\in\C$, one has
$$\dA(R')'=\bigcap_{W\in\W_R}\dA(W)''=:\M(R),$$
and the net $\M$ satisfies locality.} 
\end{quote}
{\bf Proof.} 
We first show that the net $(\dA(R')')_{R\in\C}$ satisfies locality.
This immediately follows from the fact remarked above
that if $R$ and $S$ are spacelike separated regions
in $\C$, there is a wedge $W\in\W$ with $R\subset W$ and $S\subset
W'$. For such a constellation one has
$$\dA(R')'\subset\dA(W')'=\dA(W)''\subset\dA(S')'',$$
which is the stated locality for the net $(\dA(R')')_{R\in\C}$. 

One proves in the same way that the net
$\M$ satisfies locality with respect to $\dA$, i.e.,
$\M(R)\subset\dA(R')'$ for all $R\in\C$. 
On the other hand,
$$\dA(R')'\subset\bigcap_{W\in\W_R}\dA(W')'=\bigcap_{W\in\W_R}\dA(W)''
=\M(R)\qquad\forall\,R\in\C,$$
and this completes the proof. 

\Halmos

\bigskip\bigskip\noindent%
So if one assumes wedge duality, the localization prescriptions
$L^\W(A)$, ${\bf L}^{\W}(A)$, and $L^\K(A)$ coincide and provide the
smallest localization region out of the above suggestions.
In what follows, wedge duality will be assumed, and for every local
observable $A\in\aloc$, we simply write
$L^\W(A)={\bf L}^{\W}(A)=L^\K(A)=:L(A)$.

Now the question arises in how far $L(A)$ can be considered as the
region where the observable $A$ can be measured. For this
interpretation to be consistent it is important that the localization
prescription $L$ satisfies {\em locality} in the sense that
observables with spacelike separated localization regions commute.
This does {\em not} follow from the locality assumption made for the
net $\dA$. To illustrate this, consider the wedge
$X:=W_1+e_1$, where $e_1$ is the unit vector in the 1-direction, and
its images $Y$ and $Z$ under rotations in the 1-2-plane by $120^\circ$
and $240^\circ$, respectively.  Assume a local observable $A$ to be
contained in $\dA(X)''$ and in $\dA(Y)''$, while another local
observable $B$ is contained in $\dA(Y)''$ and $\dA(Z)''$.  In this
case, the localization regions $L(A)$ and $L(B)$ are spacelike with
respect to each other, but locality of the net alone is not yet
sufficient to conclude that $A$ and $B$ should commute, since not any
two of the three wedges are spacelike separated.

Actually, this simplified sketch already points towards the
sufficient and necessary condition for locality of $L$
provided by the following theorem.

\subsubsection{Theorem (nonempty-intersection
  theorem)}\label{intersection property}
\begin{quote}{\it
Assume $\dA$ to satisfy wedge duality.

The localization prescription $\aloc\ni A\mapsto L(A)$
satisfies locality if and only if
for every finite family $W_1,\dots,W_n$ of wedges and for every
causally complete and convex region $R\in\C$ with
$\bigcap_\nu\overline{W}_\nu\subset R$, one has
$$\aloc\cap\bigcap_{1\leq\nu\leq n}\dA(W_\nu)''
\subset\dA(R')'.$$}
\end{quote}
{\bf Proof.} To prove that the condition is sufficient, let
$\del{\bf B}_\eps(L(A))$ be the boundary of the open
$\eps$-neighbourhood ${\bf B}_\eps(L(A))$ of $L(A)$ for $\eps>0$,
and define 
$$\W_A:=\{W\in\W:\,\exists X\in\W:\,A\in\dA(X)'',\,\overline{X}\subset W\}.$$ 
A class of closed subsets of the 
compact space $\del{\bf B}_\eps(L(A))$ is defined by
$$\X:=\{\del{\bf B}_\eps(L(A))\cap\overline{W}:\,W\in\W_A\}.$$
$\X$ has empty intersection,
and by the Heine-Borel property, there is a finite
subclass of $\X$ whose intersection is still
empty, i.e., there are wedges $W_1,\dots,W_n\in\W_A$ such that
$$\del{\bf B}_\eps(L(A))\cap\bigcap_\nu\overline{W}_\nu=\emptyset.$$
Due to the convexity of $L(A)$ and of wedges it follows that the region 
$$R:=\bigcap_\nu W_\nu$$
is a subset of ${\bf B}_\eps(L(A)),$ and that $R\in\B$. By the definition of
the class $\W_A$, there are wedges $X_1,\dots,X_n$ in $W_A$ such that
$\overline{X}_\nu\subset W_\nu$ for $1\leq\nu\leq n$.

Since $R\in\B\subset\C$, one now obtains from the condition that
$$A\in\aloc\cap\bigcap_\nu\dA(X_\nu)''
\subset\dA(R')'\subset\dA({\bf B}_\eps(L(A))')',$$ 
as stated. This holds for each $\eps>0$, and evidently, the
same reasoning proves that $B\in\dA({\bf B}_\eps(L(B))')'$.

Since $L(A)$ and $L(B)$ are compact, convex, and spacelike separated,
the euclidean distance between these regions is positive, and one can
choose $\eps>0$ so small that ${\bf B}_\eps(L(A))$ and ${\bf
B}_\eps(L(B))$ still are spacelike separated. As remarked in Section
\ref{notation}, ther is a wedge $X$ such that ${\bf
B}_\eps(L(A))\subset X$ and ${\bf B}_\eps(L(B))\subset X'$.  Using
wedge duality and Lemma \ref{BoY}, one concludes $$A\in\dA\left({\bf
B}_\eps(L(A))'\right)'\subset\dA(X)'',$$ and $$B\in\dA\left({\bf
B}_\eps(L(B))'\right)'\subset\dA(X')''=\dA(X)',$$ so $AB=BA$, proving
that the condition is sufficient.

To prove that the condition is necessary, let $W_1,\dots,W_n$ be a
family of wedges, and choose an $R\in\C$ with
$\bigcap_\nu\overline{W}_\nu\subset R$. Whenever
$A\in\aloc\cap\bigcap_{\nu}\dA(W_\nu)''$ and $B\in\aloc\cap\dA(X)''$
for any $X\in\W^{R'}$, locality of $L$ implies that $AB=BA$, and one
concludes that
\begin{align*}
A&\in\bigcap_{X\in\W^{R'}}(\aloc\cap\dA(X)'')'
=\bigcap_{X\in\W^{R'}}\dA(X)'
=\bigcap_{X\in\W^{R'}}\dA(X')''\\
&=\bigcap_{X\in\W_{R}}\dA(X)''=\dA(R')',
\end{align*}
where Lemma \ref{BoY} has been used in the last step. 

\Halmos

\bigskip\bigskip\noindent%
This theorem immediately implies the following corollary.
\subsubsection{Corollary}
\begin{quote}{\it
Assume $\dA$ to satisfy wedge duality, and suppose that the localization
prescription $L$ satisfies locality. If $A$ is a local observable
and $R\in\C$ is a causally complete convex open region in $\Rd$
such that $L(A)\subset R$, then $A\in\dA(R')'$.}
\end{quote}
As the following proposition shows, the additional assumption of 
wedge duality (Assumption (F) above) is sufficient to ensure locality
of $L$.

\subsubsection{Proposition}\label{wed add suf}
\begin{quote}{\it
Assume $\dA$ to satisfy wedge duality and wedge additivity.
Then the localization prescription $L$ satisfies locality.}
\end{quote}
{\bf Proof.} 
Let $A$ and $B$ be local observables with spacelike separated localization
regions. There is a wedge $W$ such that $L(A)\subset W$ and $L(B)\subset W'$.
So as soon as one proves that this implies $A\in\dA(W)''$ and $B\in\dA(W')''$,
wedge duality implies the statement. 

To this end, we consider any
$A\in\aloc$ and any wedge $W$ whose closure is spacelike separated from $L(A)$,
and show that $A\in\dA(W)'$.
This follows from wedge additivity as soon as one has found a
double cone $P$ with the property that $W\subset W+P$ and that 
$f_{A,B}$ vanishes in $W$ for all $B\in\dA(P')'$.

So fix an $\eps>0$ such that the $\eps$-neighbourhood ${\bf B}_\eps(L(A))$
of $L(A)$ is still spacelike separated from $\overline{W}$.
As in the proof of Theorem \ref{intersection property}, we choose 
a finite number of wedges $X_1,\dots,X_n$ in the class $\W_A$ such that
$$\bigcap_\nu\overline{X}_\nu\subset{\bf B}_\eps(L(A)).$$
Now define
$$P:=(-\rho e_0+V_+)\cap(\rho e_0-V_+)$$
for some $\rho>0$ (again, $e_0$ denotes the unit vector in the 
time direction), and note that $W\subset W+P$.
Fixing $\rho>0$ sufficiently small, one can make sure that
$$W\subset\left(\bigcap_\nu(X_\nu-P)\right)'.$$ 
Choosing any $B\in\dA(P')'$, one obtains
from wedge duality that the commutator function $f_{A,B}$ defined
above vanishes in the region
$$R:=\bigcup_\nu(X_\nu-P)',$$ which is a 
union of wedges. 

As in the proof of Theorem \ref{Landau}, $f_{A,B}$ can be 
shown to vanish in a larger region $N\supset R$ which is a Jost-Lehmann-Dyson
region. This can be shown by mimicking the corresponding part of that
proof, as it does not depend on the assumption that the intersection of
the closed wedges under consideration is empty. So one can
keep $A$, $B$, and the double cone $P$,
choose some double cone $\O$ with $A\in\dA(\O)$, replace
$X_1,\dots,X_n$ by $W_1,\dots,W_n$, and proceed 
like above to construct $N$.

A mass hyperboloid $H$ is
admissible with respect to $N$ only if it is admissible with respect to $R$,
and as $R$ is a union of closed wedges,
this is the case only if the whole unique shift
of the open light cone which contains $H$ is disjoint from $R$. 
But by Theorem \ref{JLD},
this implies that in particular, $f_{A,B}$ vanishes in the region $W$,
completing the proof.

\Halmos

\bigskip\bigskip\noindent%
Thomas and Wichmann have obtained a similar result for 1+3 dimensions
from slightly stronger assumptions (Theorem 4.10 in \cite{TW97a}).

One may ask what is the difference between the intersection condition
found in Theorem \ref{intersection property} and the \lq brute
force\rq\, condition that $\dA(\O)\cap\dA(P)=\dA(\O\cap P)$ for all
$\O,P\in\C$. Clearly, this condition is the stronger one of the two,
and it would imply locality of $L$ in a straightforward fashion.
Furthermore the property appears so natural that one could expect it
to be a general feature of local nets.

In \cite{Lan74}, Landau has given examples of theories which 
exhibit wedge duality, wedge additivity, and, hence,
locality of $L$ and the equivalent condition given above, while 
$\dA(\O)\cap\dA(P)\not=\dA(\O\cap P)$ for $\O,P\in\C$.

To illustrate the geometrical trick of Landau's example, start from
some local net $\dB$ of observables in 1+(s+1) dimensions, and with
every double cone $\O=(a+V_+)\cap(b+V_-)$ in $\Rd$, associate the
algebra $$\dB_0(\O):=\dB((a+\widehat V_+)\cap(b+\widehat
V_-))=:\dB(\widehat\O),$$ where, as before, $\widehat V_+$ and
$\widehat V_-$ denote the 1+(s+1)-dimensional forward and backward
light cone, respectively.

One easily checks that $\dB_0(\O)\cap\dB_0(P)$ might not coincide with
$\dB_0(\O\cap P)$, since the intersection of the 1+(s+1)-dimensional
Asgeirsson hulls of $\O$ and $P$ differs from the 1+(s+1)-dimensional
Asgeirsson hull of the intersection $\O\cap P$, i.e.,
$\widehat\O\cap\widehat P\not=\widehat{\O\cap P}$. Indeed, Landau has
given examples for theories where the corresponding algebras
differ. In particular, they differ if the \lq large\rq\, net $\dB$ has the
intersection property, i.e., if $\dB(\O)\cap\dB(P)=\dB(\O\cap P)$ for
all $\O,P\in\B$. This shows that the intersection property cannot be a
general property of all local nets of observables.

While Landau's examples do satisfy all of our above conditions, they
illustrate that the sufficient and necessary condition for $L$ to be
local is not self-evident, as it is similar to the intersection
property violated by Landau's examples. On the other hand, the fact
that all our sufficient conditions for the locality of $L$ hold,
gives some hope that locality of $L$ is a rather natural property of 
local nets.

\section{Conclusion}\label{conclusion}

Generalizing Landau's result that the algebras associated with two
strictly disjoint double cones have a trivial intersection, the {\em
empty-intersection theorem} makes it possible to associate a nonempty
causally complete, convex and compact {\em localization region} with
every {\em single} local operator of a local net. If one makes the
additional assumption of wedge duality, there is a natural way how to
obtain a smallest localization region from the empty-intersection
theorem. Even in this situation it is a nontrivial issue whether
observables with spacelike separated localization regions commute. As
a necessary and sufficient condition for this, the {\em
nonempty-intersection theorem} establishes a special intersection
property, and sufficient for this property is the additional condition
of wedge additivity, a property typically shared by models arising
from Wightman fields. As these results depend on very weak 
additional assumptions, locality of the localization prescription
$L$ turns out to be a rather natural property of local nets.

The question what the intersection of two algebras of local
observables contains has arised earlier, as, e.g., the remarks in
Section III.4.2 of Haag's monograph \cite{Haa92} show.  Haag's \lq
Tentative Postulate\rq\, that the map $\O\mapsto\dA(\O)$ be a
homomorphism from the orthocomplemented lattice of all causally
complete regions (which, in general, are neither bounded nor convex)
of Minkowski space into the orthocomplemented lattice of von Neumann
algebras on a Hilbert space does not hold in general as it stands
(cf. also Haag's heuristic remarks which illustrate the physical
limits of the postulate). But if a net satisfies wedge duality and
strong additivity for wedges, the above results, indeed, imply parts
of Haag's conjecture: for arbitrary finite families of wedges, one
obtains relations in the spirit of (III.4.7) through (III.4.11) in
\cite{Haa92} for the dual net.

The results of this article have been used for the analysis of the
Unruh effect and related symmetries of quantum fields
\cite{Kuc98,Kuc99}. Proceeding, so to 
speak, in the converse direction, Thomas and Wichmann have
investigated the implications that the symmetries providing the Unruh
effect exert on the localization behaviour of local
observables. Assuming the theory to exhibit the Unruh-effect and a
couple of (standard) technical properties including wedge additivity,
they found that the localization region of an observable $A$ with
respect to a minimal Poincar\'e covariant local net generated by $A$
is the smallest region $\O_A$ in $\B$ with the property that for every
$(a,\Lambda)\in\P_+^{\uparrow}$, one has $a+\Lambda\O_A\subset\O_A'$
if and only if $[A,U(a,\Lambda)AU(a,\lambda)^*]=0$ \cite{TW97a}.  This
definition of a localization region no longer explicitly refers to any other
operators of the net (while it does refer to the representation, which, by
the Bisognano-Wichmann symmetries, is closely related to the net). While
this interesting conclusion has been derived from the Unruh effect and
other assumptions of relevance in the above discussion, all these
assumptions have been avoided above since they are a goal rather than
a starting point of the above analysis. In this sense, the results of
Thomas and Wichmann are complementary to the above results.

\begin{appendix}
\section*{Appendix}
For the reader's convenience we include a proof that the Reeh-Schlieder
property entails weak additivity (cf. also Lemma 2.6 in \cite{TW97a}).

\subsection*{Lemma}\label {Lemma C}

\begin{quote}{\it
Let $\dA$ be a local net of local observables satisfying Conditions (A) 
through (D) above, let $\O\subset\Rd$ be a bounded open region, and let  
$a\in\Rd$ be some timelike vector. Then}
$$\C_{\O,a}:=\left(\bigcup_{t\in\reals}\dA(\O+ta)\right)''=\B(\H).$$
\end{quote}
{\bf Proof.} For any $a$ and $\O$ as above, 
let $A$ be any local observable commuting with all elements of
$\C_{\O,a}$, and pick a $B\in\dA(\O)$.
Define $f_+(t):=\langle\gO,A^*U(ta)B\gO\rangle$ and
$f_-(t):=\langle\gO,BU(-ta)A^*\gO\rangle$.
By the spectral theorem and the spectrum condition,
the Fourier transforms of these functions are (not necessarily
positive, but bounded) measures one
of which has its support in the closed positive half axis, while the 
other one has its support in the closed negative half axis. Since
$f_+$ and $f_-$ coincide by construction, it follows 
that the Fourier transform of $f_+$ (and of $f_-$)
is a measure with support $\{0\}$, i.e., some multiple of the
Dirac measure, so that $f_+$ is a constant function. Using this, the 
spectral theorem, and uniqueness of the vacuum, one concludes
$$\langle A\gO,B\gO\rangle=f_+(0)=f_+(t)=
\langle\gO,A\gO\rangle\langle\gO,B\gO\rangle
=:\overline{\alpha}\langle\gO,B\gO\rangle=\langle\alpha\gO,B\gO\rangle.$$
Since by the Reeh-Schlieder property,
$B\gO$ runs through a dense set, one concludes $A\gO=\alpha\gO$, and
since $A$ is a local observable, one obtains $A=\alpha\,\id$
since $\gO$ is cyclic with respect to $\dA(\O')$, which implies that it is
separating with respect to $\dA(\O)$ dut to locality. This proves the lemma.

\Halmos
\end{appendix}

\section*{Acknowledgements}
It was a very important help that D. Arlt, W. Kunhardt, and N. P. Landsman 
read preliminary versions of the manuscript very carefully.
I would also like to thank K. Fredenhagen for helpful discussions
and his encouraging interest. During a visit to G\"ottingen I 
obtained helpful hints from M. Lutz and M. Requardt.

The above research is part of a project that has been re-initiated at the 
Erwin Schr\"odinger Institute in Vienna, where I had the occasion to
join the project \lq Local Quantum Physics\rq\, in the autumn of 1997,
and a \lq Nachlese\rq\, meeting in the spring of 1999. I would like
to thank the ESI for the kind invitations and the warm hospitality
exhibited to me. At the ESI I profited from discussions with H.-J.
Borchers, D. Guido, S. Trebels, and E. Wichmann.

This work has been supported by the Deutsche Forschungsgemeinschaft,
the European Union's TMR Network \lq Noncommutative Geometry\rq, 
a Feodor-Lynen grant of the Alexander von Humboldt 
foundation, a part of which has been funded by the University of
Amsterdam, and a Casimir-Ziegler award of the Nordrhein-Westf\"alische
Akademie der Wissenschaften.

\end{document}